\documentclass[journal]{IEEEtran}
\usepackage[utf8]{inputenc}
\usepackage{graphicx}   
\usepackage{float} 
\usepackage{wrapfig} 
\usepackage[margin=1.0in]{geometry}
\usepackage[usenames,dvipsnames]{color}
\usepackage[breaklinks=true]{hyperref}
\hypersetup{
  colorlinks   = true, 
  urlcolor     = blue, 
  linkcolor    = blue, 
  citecolor   =  red 
}
\usepackage[capitalise]{cleveref} 
\crefformat{section}{Sec.~#2#1#3} 
\Crefformat{section}{Section~#2#1#3}
\crefformat{figure}{Fig.~#2#1#3}
\Crefformat{figure}{Figure~#2#1#3}

\definecolor{mscolor}{rgb}{0,0.5,0.5}

\title{Building a Quantum Engineering Undergraduate Program}

\author{Abraham Asfaw, Alexandre Blais, Kenneth R. Brown, Jonathan Candelaria, Christopher Cantwell, Lincoln D. Carr, Joshua Combes, Dripto M. Debroy, John M. Donohue, Sophia E. Economou, Emily Edwards, Michael F. J. Fox, Steven M. Girvin, Alan Ho, Hilary M. Hurst, Zubin Jacob, Blake R. Johnson, Ezekiel Johnston-Halperin, Robert Joynt, Eliot Kapit, Judith Klein-Seetharaman, Martin Laforest, H. J. Lewandowski, Theresa W. Lynn, Corey Rae H. McRae, Celia Merzbacher, Spyridon Michalakis, Prineha Narang, William D. Oliver, Jens Palsberg, David P. Pappas, Michael G. Raymer, David J. Reilly, Mark Saffman, Thomas A. Searles, Jeffrey H. Shapiro, and Chandralekha Singh
\thanks{(Received, revised, accepted dates of manuscript.) This work was partially supported by the U.S. National Science Foundation under grant EEC-2110432.  S.~Girvin, T. A. Searles, and S. E. Economou were supported by the U.S. Department of Energy, Office of Science, National Quantum Information Science Research Centers, Co-design Center for Quantum Advantage (C2QA) under contract number DE-SC0012704. M.~Raymer acknowledges support from the NSF Engineering Research Center for Quantum Networks (CQN), led by the University of Arizona under NSF grant number 1941583. E. Johnston-Halperin acknowledges support from NSF C-ACCEL 2040581. H. J. Lewandowski acknowledges support from NSF QLCI OMA–2016244.  L.~D.~Carr, H.~M.~Hurst, E.~Kapit, and T.~Lynn acknowledge support from NSF QLCI-CG OMA-1936835. C.R.H.~McRae and D.P.~Pappas were supported by NIST NQI and QIS efforts, as well as the U.S. Department of Energy, Office of Science, National Quantum Information Science Research Centers, Superconducting Quantum Materials and Systems Center (SQMS) under the contract No. DE-AC02-07CH11359. A. Blais was supported by the Canada First Research Excellence Fund. S.~Michalakis was supported by Caltech's Institute for Quantum Information and Matter (IQIM), a National Science Foundation (NSF) Physics Frontiers Center (NSF Grant PHY-1733907). M. Saffman was supported by NSF QLCI-HQAN Award 2016136. This article was presented in part at the Quantum Undergraduate Education \& Scientific Training (QUEST) Workshop and at the SPIE Photonics for Quantum Symposium. \textit{(All authors contributed equally to this work; authorship is listed alphabetically by last name.) (Corresponding author: Lincoln D. Carr, lcarr@mines.edu)}}
\thanks{}
\thanks{A. Asfaw and B. R. Johnson are with IBM Quantum, IBM T. J. Watson Research Center, Yorktown Heights, NY, U.S.A.}
\thanks{A. Blais is with Institut Quantique and D\'epartement de Physique, Universit\'e de Sherbrooke, Sherbrooke J1K 2R1 QC, Canada and Canadian Institute for Advanced Research, Toronto M5G 1M1 ON, Canada}
\thanks{K. R. Brown and D. Debroy are with the Duke Quantum Center, Department of Physics, Department of Electrical and Computer Engineering, and Department of Chemistry, Duke University, Durham, NC 27708, U.S.A.}
\thanks{J. Candelaria is with SystemX Program and Department of Electrical Engineering, Stanford University, Stanford, CA 94305, U.S.A.}
\thanks{C. Cantwell is with Department of Physics and Astronomy, University of Southern California, Los Angeles, California 90089, U.S.A.}
\thanks{L. D. Carr and E. Kapit are with the Quantum Engineering Program and Department of Physics, Colorado School of Mines, Golden, CO 80401, U.S.A.}
\thanks{J. Combes is with the Department of Electrical, Computer and Energy Engineering, University of Colorado Boulder, Boulder, Colorado 80309, U.S.A.}
\thanks{J. M Donohue is with the Institute for Quantum Computing, University of Waterloo, Waterloo, ON N2L 3G1 Canada}
\thanks{S. E. Economou is with the Department of Physics, Virginia Tech, Blacksburg, Virginia 24061, U.S.A.}
\thanks{E. Edwards is with the IQUIST, University of Illinois Urbana-Champaign, Urbana, IL 61801, U.S.A.}
\thanks{M. F. J. Fox and H. Lewandowski are with JILA, National Institute of Standards and Technology and the University of Colorado, Boulder, CO 80309, U.S.A. and Department of Physics, University of Colorado Boulder, Boulder, CO 80309, U.S.A.}
\thanks{M. F. J. Fox is with the Department of Physics, Imperial College London, Prince Consort Road, London, SW7 2AZ, UK}
\thanks{S. Girvin and D. Debroy are with the Yale Quantum Institute and Department of Physics, Yale University, New Haven, CT 06520, U.S.A.}
\thanks{A. Ho is with Google Research, Venice, CA 90291, U.S.A.}
\thanks{H. M. Hurst is with the Department of Physics \& Astronomy, San Jos\'{e} State University, San Jos\'{e}, California 95192, U.S.A.}
\thanks{Z. Jacob is with the School of Electrical and Computer Engineering, Purdue University, West Lafayatte, Indiana - 47907, U.S.A.}
\thanks{E. Johnston-Halperin is with the Department of Physics, The Ohio State University, Columbus, OH 43210, U.S.A.}
\thanks{R. Joynt and M. Saffman are with the Department of Physics, University of Wisconsin-Madison, 1150 University Avenue, Madison, WI 53706, U.S.A.}
\thanks{J. Kelin-Seetharaman is with the Quantitative Biosciences and Engineering Program and Department of Chemistry, Colorado School of Mines, Golden, CO 80401, U.S.A.}
\thanks{M. Laforest is with ISARA Corporation, Waterloo, Ontario N2L 0A9, Canada}
\thanks{T. W. Lynn is with the Department of Physics, Harvey Mudd College, Claremont, CA 91711, U.S.A.}
\thanks{C. R. H. McRae is with the Department of Physics, University of Colorado Boulder, Boulder, CO 80309, U.S.A., and National Institute of Standards and Technology Boulder, Boulder, CO 80305, U.S.A.}
\thanks{C. Merzbacher is with SRI International, Boulder, CO 80302}
\thanks{S. Michalakais is with the Institute for Quantum Information and Matter, California Institute of Technology, Pasadena, CA 91125, U.S.A.}
\thanks{P. Narang is with the John A. Paulson School of Engineering and Applied Sciences, Harvard University, Cambridge, Massachusetts 02138, U.S.A.}
\thanks{W. D. Oliver is with the Department of Electrical Engineering and Computer Science and Lincoln Laboratory, Massachusetts Institute of Technology, Cambridge, MA 02139, U.S.A.}
\thanks{J. Palsberg is with the Department of Computer Science, University of California – Los Angeles, Los Angeles, California 90095, U.S.A.}
\thanks{D. P. Pappas is with the National Institute of Standards and Technology, Boulder, CO 80303, U.S.A.}
\thanks{M G. Raymer is with the Department of Physics and Oregon Center for Optical, Molecular and Quantum Science, University of Oregon, Eugene, OR 97403, U.S.A.}
\thanks{D. J. Reilly is with the ARC Centre of Excellence for Engineered Quantum Systems, School of Physics, The University of Sydney, Sydney, NSW 2006, Australia and Microsoft Quantum Sydney, The University of Sydney, Sydney, NSW 2006, Australia}
\thanks{T. A. Searles is with the IBM-HBCU Quantum Center, Department of Physics \& Astronomy, Howard University, Washington, DC 20059, U.S.A.}
\thanks{J. H. Shapiro is with the Department of Electrical Engineering and Computer Science, Massachusetts Institute of Technology, Cambridge, MA 02139, U.S.A.}
\thanks{C. Singh is with the Department of Physics and Astronomy, University of Pittsburgh, Pittsburgh, PA 15260, U.S.A.}
}

\date{June 2021}

\begin{document}

\maketitle

\begin{abstract}
The rapidly growing quantum information science and engineering (QISE) industry will require both quantum-aware and quantum-proficient engineers at the bachelor's level. We provide a roadmap for building a quantum engineering education program to satisfy this need. For quantum-aware engineers, we describe how to design a first quantum engineering course accessible to all STEM students.  For the education and training of quantum-proficient engineers, we detail both a quantum engineering minor accessible to all STEM majors, and a quantum track directly integrated into individual engineering majors.  We propose that such programs typically require only three or four newly developed courses that complement existing engineering and science classes available on most larger campuses.  We describe a conceptual quantum information science course for implementation at any post-secondary institution, including community colleges and military schools. QISE presents extraordinary opportunities to work towards rectifying issues of inclusivity and equity that continue to be pervasive within engineering. We present a plan to do so and describe how quantum engineering education presents an excellent set of education research opportunities. Finally, we outline a hands-on training plan on quantum hardware, a key component of any quantum engineering program, with a variety of technologies including optics, atoms and ions, cryogenic and solid-state technologies, nanofabrication, and control and readout electronics. Our recommendations provide a flexible framework that can be tailored for academic institutions ranging from teaching and undergraduate-focused two- and four-year colleges to research-intensive universities.
\end{abstract}

\tableofcontents

\section{Introduction to Quantum Engineering}
\label{sec:introduction}

Quantum information science combines our understanding of nature at its most fundamental level---quantum mechanics ---with information theory. From an applications standpoint, advancements in this field now rely on incorporating an engineering approach to better design, integrate, and scale quantum technologies.
For example, the landmark quantum advantage result achieved in 2019~\cite{arute2019quantum}, in which a quantum computer met a computational benchmark not achievable on the same time scale with present classical computing resources, made heavy use of a range of engineering disciplines to construct a machine capable of quantum speed-up. This is one example of the many recent advances in quantum information science (QIS) spanning algorithms, architectures, and qubit technologies, including atoms and ions, semiconductors, superconductors, as well as supporting hardware in integrated optics, and microwave and RF control and readout~\cite{bruzewicz2019trapped,zhang2019semiconductor,kjaergaard2020superconducting,blais2020quantum}.  This new chapter of discovery and innovation is centered on novel devices that employ non-classical states, superposition, and entanglement to create technological advantage over classical systems. In addition, devices based on these aspects of quantum physics could serve as a foundational technology, similar to the role of semiconductors in the 20th century. In doing so, QIS is predicted to open up otherwise impossible vistas in communication, computation, and sensing.  Future engineers are needed to address grand challenges such as scalability and identifying unique real-world opportunities for quantum systems, and indeed industry positions reflect this need~\cite{merzbacher2021,hughes2021}.
Given this expectation, incorporating quantum information into formal engineering curricula will prepare students to work at the forefront of current science and technology, and drive future growth across multiple engineering sectors.

So, what is this new field of \emph{quantum engineering} and what are the implications for education programs?  As defined by the U.S. National Quantum Initiative Act, quantum information science is ``the use of the laws of quantum physics for the storage, transmission, manipulation, computing, or measurement of information''~\cite{raymer2019us}. Building on this definition, we define quantum engineering as the application of engineering methods and principles to quantum information systems and problems. This includes the work of both quantum-aware engineers and quantum-proficient engineers-- and is necessarily redefining the field to be quantum information science and engineering (QISE).

Increasing the engineering talent flow into QISE could vastly accelerate the development of quantum technologies, some of which might be harnessed to tackle some of the world's most pressing problems, such as more efficient nitrogen fixation~\cite{reiher2017elucidating}, making artificial light-harvesting photosynthetic complexes for clean energy~\cite{ball2018photosynthesis}, and addressing the rapidly approaching end of Moore's law and subsequent limitation on computing resources~\cite{theis2017end}.  It is also a major opportunity to broaden participation in terms of diversity, equity, and inclusion.  In this regard, it has been argued that modifications and innovations in the engineering portion of that pipeline, as well across the physical and computational sciences---perhaps spanning kindergarten to Ph.D. with many on-ramps and opportunities along the way---will be vital for developing the workforce necessary to capitalize on the promise of QISE.
There is a pressing educational gap between, on the one hand, excitement generated both by the popular media and the increased interest in introducing QISE in secondary school, and on the other hand, quantum-related graduate programs focused mainly on PhDs, with a few MS programs as well. This undergraduate gap can be addressed in the near term, and closing it will likely have substantial impact on the quantum workforce. Finally, many engineers traditionally do not pursue a PhD, but rather a Bachelor's or at most a Master's. Specializing in quantum engineering will offer a similar educational pathway to professional life as other traditional engineering disciplines with specializations, such as bioengineering. Thus, within the context of emerging quantum education activities at the K-12, Masters, and workforce upskilling levels, this article focuses its attention on the existing gap in the quantum engineering pipeline at the \emph{undergraduate} level.

Within undergraduate programs, quantum mechanics has long been taught in physics departments, but QISE is cross-disciplinary and demands a workforce that draws from formal education programs in departments including applied mathematics, chemistry, computer science, electrical engineering, materials engineering, and molecular engineering, to name a few. Well-established Ph.D. programs and undergraduate physics programs must now be complemented by new efforts that broaden the population of quantum-proficient and quantum-aware scientists and engineers. This need is supported by the current demand for traditional engineering skills in quantum industry, which is well documented~\cite{Fox2020}. Thus, one of the tensions in discussing new formal education programs in QISE surrounds sufficiently training students in quantum engineering, while not overly diluting their broader engineering degree so that students retain a solid foundation for many decades of continued learning and professional experience. We need to prepare T-shaped engineers, who have deep core knowledge and skills in some engineering discipline, while also having breadth so that they are agile and adaptable for interdisciplinary innovation in a quickly changing technological and scientific landscape~\cite{oskam2009t}.

This article lays out a detailed initial road-map for engineering schools and departments to drive quantum engineering education forward. We focus on undergraduate quantum engineers and the recommendations are intended to be tailored for different academic contexts from teaching and undergraduate-focused four-year colleges and universities to research-intensive universities---there is no one-size-fits-all solution. In Sec.~\ref{sec:context}, we provide a brief description of the technological, educational, and logistical context for quantum engineering, which also helps define the field. To facilitate quantum-awareness among budding engineers, in Sec.~\ref{sec:firstCourse} we present different pathways to build a first course in QISE accessible to any STEM major.  To train quantum-proficient engineers, in Sec.~\ref{sec:complete_program} we describe how to create a more complete undergraduate quantum engineering program.  This includes QISE education research; a complete quantum engineering minor with three existing working examples; a quantum track within an engineering major with one current working example; and some remarks on a potential future quantum engineering major, which we believe is premature at this stage.  Then in Sec.~\ref{sec:diversity}, we discuss how QISE presents an extraordinary opportunity to have a major impact on equity and inclusion in engineering as a whole, and provide specific recommendations to do so. In Sec.~\ref{sec:hands-on}, we walk through the most common quantum technologies and sketch hands-on training programs in each of them, which can be adapted to different academic environments.  Finally, in Sec.~\ref{sec:recommendations} we summarize our recommendations.

The roadmap will also prove useful to other science departments creating their own QISE programs or partnering with engineering programs to do so.  Additionally, a key recommendation of this paper centers around an introductory ``Quantum 101" course with minimal math content, like Anthropology 101 or Psychology 101, which we feel should be implemented at as many universities and colleges as possible to provide both an entry point to the field and a route towards promoting quantum awareness among the general population. Community colleges especially have returning older students as well as many high school students taking classes, and can including a general QISE course could have an outsize impact.  Since a significant fraction of engineering students are transfers from two-year colleges~\cite{national2016barriers}, this further justifies the design and wide-spread adoption of a "Quantum 101 course."  We include Quantum 101 in Sec.~\ref{sec:complete_program} as both broader impact and a recruiting tool for quantum engineering.

\section{Undergraduate Quantum Engineering in Context}
\label{sec:context}

\subsection{Technology, Industry, and Opportunity}
\label{ssec:industry}

The modern information age rests in large part on a foundation of semiconductor materials and devices that, at a fundamental level, require a quantum mechanical description. For example, semiconductor devices are the driving force behind modern computing, sensing, and networking technologies. In addition, an understanding of quantum mechanical band structure has led to the development of transistors, lasers, and photodetectors, which transmit and receive the glut of data carried by the internet's fiber-optic backbone. Today, the continued miniaturization of semiconductor devices is approaching the end of Moore's law. As the transistor size standard approaches the semiconductor atomic lattice scale,  quantum effects begin to limit rather than enable further progress.  But, as has happened before in engineering, these limitations have spurred progress as we seek to turn a bug into a feature, and develop new quantum-enhanced, rather than quantum-limited, technologies.

Quantum computers, whose non-classical properties seem to evade the bounds of the extended Church-Turing hypothesis, could provide breakthrough capabilities in optimization~\cite{Boixo2014} and machine learning~\cite{biamonte2017quantum,huang2021power}.  They also have the unrivaled ability to efficiently simulate quantum systems~\cite{Georgescu2014,altman2021quantum} since they are quantum systems themselves, and thus will find uses in chemistry, molecular biology, materials science, and drug discovery, to name a few applications.  Quantum communication will enable quantum computers to be connected in a quantum internet~\cite{Kimble2008,Wehner2018}, which will open up a host of possibilities, such as quantum secret sharing.  Quantum sensing will open up new measurement modalities and sensitivities, such as Quantum Positioning Systems (QPS) capable of autonomous navigation (GPS-free) with accuracy down to the centimeter level.

Many of the predicted technology implications and corresponding investment can be traced back to quantum key distribution (QKD), which was proposed by Bennett and Brassard in 1984~\cite{Bennett1984}. This application allows two communicating parties to establish a secret key for secure communication in the presence of an all-powerful eavesdropper. QKD's potential importance was magnified enormously in 1994, when Peter Shor published a quantum-computer algorithm~\cite{Shor1994} for breaking the Rivest-Shamir-Adelman (RSA) public-key infrastructure on which internet commerce currently depends. The possible vulnerability of RSA soon spurred research---both theoretical and experimental---in quantum communication and quantum computing that went far beyond QKD and algorithmic attacks on cryptographic protocols like RSA.

Currently, one of the largest areas of activity in QISE is associated with quantum computers~\cite{alexeev2021quantum} and the creation of quantum networks~\cite{awschalom2021development}. There are already over 300 quantum simulators, or specialized ``analog'' quantum computers worldwide, built on over 10 distinct physical architectures~\cite{altman2021quantum}.  More general ``digital'' quantum computers already report achieving a quantum advantage~\cite{arute2019quantum,zhong2020quantum,alexeev2021quantum}, that is, meeting computational benchmarks not possible on a reasonable time scale with present classical computing resources.  There is also ample activity related to the future of quantum sensing~\cite{Caves1981,Bondurant1984,Xiao1987,LIGO2013}. Quantum sensing has been applied in magnetometry~\cite{Maze2008} and ultra-precise clocks~\cite{Bloom2014}, among other areas. There may also be new untapped possibilities for improved classical information transmission using quantum communication~\cite{Holevo1998,Hausladen1996,Schumacher1997,Giovannetti2004}. Moreover, quantum communication and sensing principles have led to proposals for improving the angular resolution of astronomical imagers~\cite{Gottesman2012,Tsang2019} and the sensitivity of microwave radars~\cite{Barzanjeh2015,Luong2020,Barzanjeh2020, Shapiro2020}, and additional such advances are anticipated.

Due to advances in QISE coupled to its potential scientific and societal impacts, quantum research has correspondingly expanded beyond academia and government laboratories into industry. As of Fall 2019 there were already more than 87 quantum-related companies, spanning sensors, networking and communications, computing hardware, algorithms and applications, and facilitating technology, and this industry continues to expand~\cite{Fox2020}. Recent research by Fox, Zwickl, and Lewandowski~\cite{Fox2020} points out that although researchers in this field are often Ph.D. physicists, as quantum industry products are moved out of development and into production, the need for engineers will increase. This work goes on to summarize the skill set valued by employers in the quantum industry: coding, statistical methods for data analysis, laboratory experience, electronics knowledge, problem-solving, materials properties, and quantum algorithms. Inasmuch as nearly all quantum computation, communication, and sensing developments will involve a great deal of classical engineering, having some level of quantum awareness will be sufficient for many engineering graduates~\cite{NSTC2018}. Examples include microwave engineers who will work on interconnections within superconducting quantum computers, photonics engineers who will work on the fiber links for quantum networks, and control engineers who will work on the various control systems required by quantum computing technologies. This quantum awareness could potentially be achieved with a single course or a two-semester sequence, see Secs.\ref{sec:firstCourse} and~\ref{ssec:concepts}, or with an undergraduate minor or track, see Secs.~\ref{ssec:minor}-\ref{ssec:track}.

\subsection{The Quantum Education Landscape}

The rapid growth of QISE as an academic discipline and viable career path for graduates has led to a growing number of formal quantum education efforts at all academic levels. Outreach efforts to a broad range of nonspecialist audiences have also been developed in limited contexts recent years, and we expect that education and outreach efforts will continue to expand. To supplement this paper's emphasis on undergraduate quantum engineering  education, we discuss how undergraduate quantum engineering would fit into the wider quantum education landscape, with a focus on the U.S. national landscape, and how those relationships, along with lessons learned in other activities, could inform the development of undergraduate quantum engineering programs and courses.

At the graduate level, a growing number of institutions have already launched or are developing master’s degree programs, offering bachelor’s degree holders in several STEM fields the specialized education and professional training to help them transition into the quantum workforce---or in some cases into Ph.D. programs in QISE. These programs face many challenges that undergraduate quantum engineering will also encounter, such as the need to accommodate a variety of incoming technical backgrounds and prior exposure levels to quantum science; the need to recruit and support a diverse student population in order to promote equity and a broad range of creative perspectives in the field as a whole; and the need to train teaching assistants and faculty from a wide range of departments to teach interdisciplinary QISE courses.

Outside higher education, numerous K-12 and public outreach programs in quantum information already exist or are under development, but the reach of such programs remains limited and impact is relatively unknown. In an effort to begin establishing content frameworks for broadly introducing QISE into K-12 classrooms, museums and other learning environments, an NSF-sponsored workshop in 2020 drafted a set of nine Key Concepts for Future QIS Learners~\cite{keyconcepts} that can be adapted for engineering contexts. The National Q-12 Education Partnership and the Q2Work program are collaborating with teachers to expand these concepts for different ages and subjects and supporting the development of K-12 and public education initiatives~\cite{q12}. To increase quantum literacy among educators and community stakeholders, and to develop curricula, the NSF supports teacher workshops that are piloting lesson design and implementation as well as convergence accelerators QuSTEAM and the National Quantum Literacy Network~\cite{aiello2021achieving}.
Complementing these efforts are numerous summer camps, after-school programs, and online courses for students interested in QISE. These programs can provide inspiration and even a pipeline for students as they consider a future in quantum engineering.

K-12 formal quantum education is in its infancy and will require significant resources for full-scale implementation. This includes integrating quantum training into teacher professional development programs, researching and developing effective curricular models, and promoting long-term public awareness and engagement. Additionally, coordination with state and local education stakeholders is necessary for broad implementation.

As programs begin to scale up over the coming decades, they will begin to create a new population of students who enter college already primed with an interest in QISE and fuel the quantum information revolution, much as classical computer classes and opportunities fueled the classical information revolution starting in the 1980s.

The pressing need for quantum-proficient and quantum-aware engineers in the workforce, the growing set of graduate programs tailored to interests and ambitions in the field, and the expanding set of outreach and education opportunities for K-12 students currently leave a clear gap in the trajectories available to many quantum-interested students during their undergraduate years.
It is essential that academia support and develop courses that address this gap at diverse types of institutions---including community colleges, undergraduate colleges, and large and small research intensive universities---and promote a wide range of pathways into the field.

For instance, because a large fraction of engineering students nationwide transfer from two-year colleges~\cite{EnhancingCC2006}, community colleges and four-year institutions need to partner to remove barriers for students to make this transition in quantum engineering, just as in other STEM majors, tracks, and minors. To further support students as they move through their education, institutions could tie the development of quantum engineering programs to K-12 activities that connect with secondary school students and educators.
Those having experience in K-12 and outreach programs have indicated that initial exposures should be varied to engage a broad range of potential future quantum engineers. For some students, quantum games~\cite{cantwell2019quantum} could be an excellent opportunity to pique interest, build intuition, motivate more future rigorous study, and even deepen understanding through repeated practice in different learning contexts. For others, examples of practical applications, interdisciplinarity, and societal impact embedded within courses could be more compelling. Professionals working in K-12 and outreach efforts indicate that demystifying the field rather than emphasizing its exotic attributes may lower the barrier for entry, and potentially attract a larger, more diverse student population, as students become aware that QISE is more than an intellectual exercise and offers both existing technological applications and stable and wide-ranging job opportunities. Future quantum engineering programs at the undergraduate level can perhaps improve retention by providing regular examples of existing career trajectories and facilitating mentoring relationships~\cite{TEAMUP2020}. Anecdotal experience suggests that deliberate attention to these areas could be more critical in quantum engineering than in a more established field, where many students have career models available in the form of relatives and other community members.

Finally, a key component of QISE education is continuing education, called upskilling in the industrial context. As mentioned, a few universities already provide a master's program or graduate professional certificate in quantum engineering geared towards students with an existing undergraduate degree in STEM. Others offer online learning and certification for existing professionals. A growing number of  online continuing education platforms such as EdX, Coursera, etc., are offering formal quantum courses for the general public~\cite{freericks2018,franklin2021}. Informally, there have been a rapidly growing number of university courses placed on YouTube for public viewing. However, these programs and courses appear to be insufficient, or are too narrow in scope, to meet the rapidly increasing need for quantum proficiency across multiple disciplines. Given that the fields that contribute to QISE are not gender, racially, or ethnically diverse, upskilling programs are also unlikely to result in an increase in diversity across the technical workforce.

This all presents an opportunity for universities building their quantum engineering profile and programs. Introductory undergraduate courses such as those described in Sec.~\ref{ssec:concepts} and Sec.~\ref{sec:firstCourse} can be leveraged as part of a university’s MOOC program for existing workforce training.
Over and above upskilling, currently the workforce needs~\cite{merzbacher2021,hughes2021} are such that a large number of new BS/BE recipients will need to have quantum engineering education already in place.  In the following sections, we lay out a plan for accomplishing this goal.

\section{Building a First QISE Course for STEM Students}
\label{sec:firstCourse}

Designing introductory courses in QISE is challenging because the field is in rapid technological flux and because we need courses that are accessible to students from a wide variety of disciplines having varied mathematical and scientific preparation levels. In this sense, the challenge is similar to that faced by computer science in the early days.  In addition, local faculty expertise varies at different colleges and universities. An introductory course might be designed for first year college students, graduate students, or anyone in between. As a result, we have chosen to present our recommendations for an introductory QISE course as a set of modules from which a course can be built and then tailored to meet the individual needs of the students, program, and faculty; some of these modules, such as the introduction to the gate model, can be made accessible to students at any level (labeled with an E) below, whereas others like  quantum noise are likely more appropriate for advanced students (labeled with A).

This course is designed overall for engineering or at least STEM students as it has significant mathematical content; in Sec.~\ref{ssec:concepts} we suggest a separate Quantum 101 course which is accessible to non-STEM students.  We strongly recommend that a first QISE course as described in this section assume only a background in high school and freshman physics.  In contrast, the Quantum 101 course in Sec.~\ref{ssec:concepts} assumes no physics background at all. However, even for engineering students one may want to avoid continuous variable systems in introductory courses, even unentangled, single-particle ones.  Studying them in depth typically requires quite a lot of pre-requisite specialized mathematical knowledge, although some QISE educators have explored other approaches~\cite{private2021}. Some studies suggest the discrete-variable or ``spin-first'' approach to quantum mechanics provides more opportunities for students to understand the underlying concepts independently from the complex mathematical calculations often associated with quantum mechanics~\cite{Sadaghiani2016}. Appropriate to a college-level class, we assume students will have taken a linear algebra course beforehand, and recommend spending the first week on review of the basics of linear algebra as a refresher. Alternatively, one could choose not to rely on linear algebra as a pre-requisite and teach the required concepts as part of the course (at the expense of covering less ground in QISE) -- see Sec.~\ref{ssec:linearAlgebra}. We envision each module taking 1-3 weeks of a standard semester course, depending on depth and the educational level of the students.

The goal of our recommendation is to be able to easily combine modules to create an introductory course at any level based on the goals of the program. To use a few of the authors of this paper as examples, Girvin has taught an introductory quantum computing course aimed at first-year students onward that roughly followed the sequence of modules 1-2-3-4-6; Kapit taught an advanced course aimed at preparing seniors and graduate students for further specialized courses and research, which proceeded as 2-3-8-9-10-4; Blais taught similar level students with the sequence 2-3-7-4-6-11; Economou teaches two courses, one purely on quantum software, roughly following 1-2-3-7-4-6, and one focused on physical platforms and their control, following 8-9-10-11; and Lynn has taught an introductory course aimed at sophomores onward but sometimes taken by first-year students and even advanced high-school students, which followed the rough sequence 2-3-11-4-6-7, interspersing ideas from 1.  Carr has used module 0 as an add-on to a variety of STEM classes with quantum content when students have little to no knowledge of linear algebra.
The choice of modules would also be informed by the selection of more advanced courses the program offers beyond this introduction. For example, a program that offers a dedicated quantum algorithms class might de-emphasize much of module 4 in the introductory course, and a program that offers microwave engineering courses would certainly want to include module 9.  Such a course could be titled, ``Introduction to Quantum Engineering.''

Comprehensive learning goals may vary somewhat from course to course and should be set for any implementation of the modules. The Key Concepts for Future Quantum Information Science Learners~\cite{keyconcepts} presents a set of essential QISE ideas, but learning goals based on this must be developed. It is an open engineering education research question as to what learning goals would be appropriate in the context of undergraduate quantum information engineering. Finally, we note that although we have not included a survey of pre-quantum information era quantum mechanics, sometimes called ``Quantum 1.0'', explicitly in our modules, it is implied in many of our topics as an appropriate background to ``Quantum 2.0'', i.e., QISE.  Alternately, Quantum 1.0 can be included as a separate module on extant pre-QISE technologies such as lasers, MRIs, atomic clocks for navigation, the photoelectric effect in motion sensors, etc., for example as a non-mathematical Module 0 in place of linear algebra.

The modules below and other course recommendations are focused more on quantum computing, with less content covering quantum communication and quantum sensing. We support and encourage the development of additional modules in these critical QISE pillars, which will be essential for providing a complete curriculum.

\subsection{Module 0: Linear Algebra for QISE (E)}
\label{ssec:linearAlgebra}
Vector spaces (superposition, concept of a basis), linear transformations, matrix multiplication, non-commutativity, diagonalization, inversion, Hermitian and unitary operators, trace and partial trace, outer and tensor products, scaling up to larger matrices numerically.  These concepts can be introduced in the context of the single qubit, i.e. $2\times 2$ matrices, and their tensor products.  As linear algebra is a strong prerequisite for most QISE, students can either take it as a separate course or it can be included here as a focused unit.  The alternate option is a non-mathematical quantum concepts course as in Sec.~\ref{ssec:concepts}.

\subsection{Module 1: Classical information theory (E)}
Basics of bits, gates, communication, randomness and statistics, error correction, parity and data compression. Discusses the basics of computation itself,  shows that the number of distinct programs mapping $n$ input bits to $m$ output bits is doubly exponential $N=2^{m2^n}$ and introduces the notion of a universal classical gate set that can reproduce any of this enormous number of programs. Vector spaces could be introduced at this point, as in Mermin's book~\cite{Mermin-book}, where classical bits are represented as vectors and gates as matrices. This eases the transition to qubits.

\subsection{Module 2: One and two quantum bits (E)}
Quantum bits, superposition states, measurements and the Born rule. Single-qubit Hilbert space: linear operators, Dirac notation, orthonormal bases and basis changes, qubit rotations, and the Bloch sphere.  Expectation values and variance of measurement results.  This introduces students to the basics of quantum theory in the concrete context of the analytically solvable problem of 1 or 2 qubits.  It helps students understand multi-qubit Hilbert space and operators, leading to tensor product spaces and the combinatorial complexity explosion.

\subsection{Module 3: Two-qubit gates and entanglement (E) }
The CNOT gate and the circuit model of computation. Bell states and non-classical correlations. A typical example would be the spin singlet or Bell states, in which information is encoded in the global system while no information is contained in the constituent qubits. Quantum dense coding and monogamy of entanglement. The no-cloning theorem and state teleportation. Universal quantum gate sets. This shows students how to build quantum computation from basic elements (gates) and some of the surprising outcomes.  Depending on the programmatic emphasis, this module could focus on entanglement more generally, for instance, in single-ion optical atomic clocks, cold molecular ion spectroscopy, quantum communications, random key generation, etc.

\subsection{Module 4: Quantum algorithms (E/A)}
Early examples of quantum advantage in computation: the Deutsch, Deutsch-Jozsa, Bernstein-Vazirani, Simon's algorithms. Phase kick back from controlled unitaries. Oracle algorithms. Grover's algorithm, phase estimation, and the quantum Fourier transform. In discussing Shor's algorithm, one may want to focus on the QFT and the period-finding algorithm more than factoring itself, since the factoring application depends on number theoretic results which are less relevant to quantum algorithms more broadly.  Students will gain an understanding of the breadth of quantum algorithm development.

\subsection{Module 5: NISQ devices and algorithms (E/A)}
Noisy intermediate scale quantum devices. Heuristic algorithms, including the Variational Quantum Eigensolver (VQE), the Quantum Approximate Optimization Algorithm (QAOA), and possibly machine learning algorithms. Error mitigation. Use of online software and cloud-accessible hardware.  Students will understand how to program actual quantum hardware and learn some minimal quantum software skills.

\subsection{Module 6: Quantum error correction (E/A)}
Quantum computers are analog and errors are continuous, but \emph{measured} errors are discrete. Error models: coherent errors, incoherent errors as coherent errors in a larger Hilbert space, correlated errors. Repetition code for dephasing or bit-flip errors. Shor code. Concatenation, code capacity threshold vs.\ fault-tolerance threshold.  Error correction is one of the most essential topics in QISE and students need a careful introduction to clarify the contrast with much easier error correction methods on classical computers.

\subsection{Module 7: Quantum communication and encryption (E)}
Inability to perfectly distinguish non-orthogonal states and no-cloning theorem. The BB84 quantum key distribution protocol. Entanglement-based quantum key distribution protocol (E91). Entanglement swapping and quantum repeater networks. Error correction in communication and entanglement distillation. This module can build on module 3, and is the intro to the power of QISE for communications.

\subsection{Module 8: Hamiltonians and time evolution (A)}
Eigenstates and eigenenergies of a Hamiltonian. The Schr\"odinger equation and time evolution. Expectation values; motion and transitions as interference phenomena. The harmonic oscillator and general $N$-level systems. Basic properties of systems with multiple identical particles. This module is especially useful for building a knowledge of quantum and classical control systems, since quantum gates are based on the underlying dynamics.

\subsection{Module 9: Dynamics with time-varying Hamiltonians (A)}
Dynamics and control of two-level systems subject to AC fields. Quantum mechanics in a rotating frame, and the rotating wave approximation. Rabi oscillations. Control of harmonic systems with an auxiliary anharmonic element. The quantum adiabatic theorem.  This module naturally builds on module 8, or can be expanded and substituted in place of it.  AC fields such as microwaves are key to classical control of quantum systems.

\subsection{Module 10: Open quantum systems (A)}
The density matrix formulation of quantum mechanics. Entangling and non-entangling noise. Fermi's Golden Rule. Models for a bath. Reduced density matrices. Physical noise mechanisms. Quantifying coherence through estimates of relaxation and dephasing times.  An essential concept in QISE is the fragility of quantum states.  This module can provide underlying knowledge to comprehend the severity of the decoherence problem.

\subsection{Module 11: Physical quantum bits (E/A)}
Broad overview of candidate systems for quantum computing. Superconducting qubits, trapped ions, spin qubits. At a more advanced level, one can also present photonic systems, neutral atoms and topological qubits. One could also formulate this module as a more in-depth exploration of a single class of qubits.

\subsection{Module 12: Quantum Sensing Modalities (E/A)}
Quantum-enhanced resolution in optical interferometry:  classical operation versus N00N-state (entangled) operation.  Heisenberg uncertainty principle for the quantum harmonic oscillator:  coherent states and squeezed states.  This will demonstrate to students the basic principles of quantum-enhanced accuracy in optical interferometry, including  coherent-state operation with standard-quantum-limit scaling versus squeezed-state operation with Heisenberg scaling.

\section{Creating a Complete Undergraduate Quantum Engineering Program}
\label{sec:complete_program}

\begin{figure*}
    \begin{center}
        \includegraphics[width=0.8\textwidth]{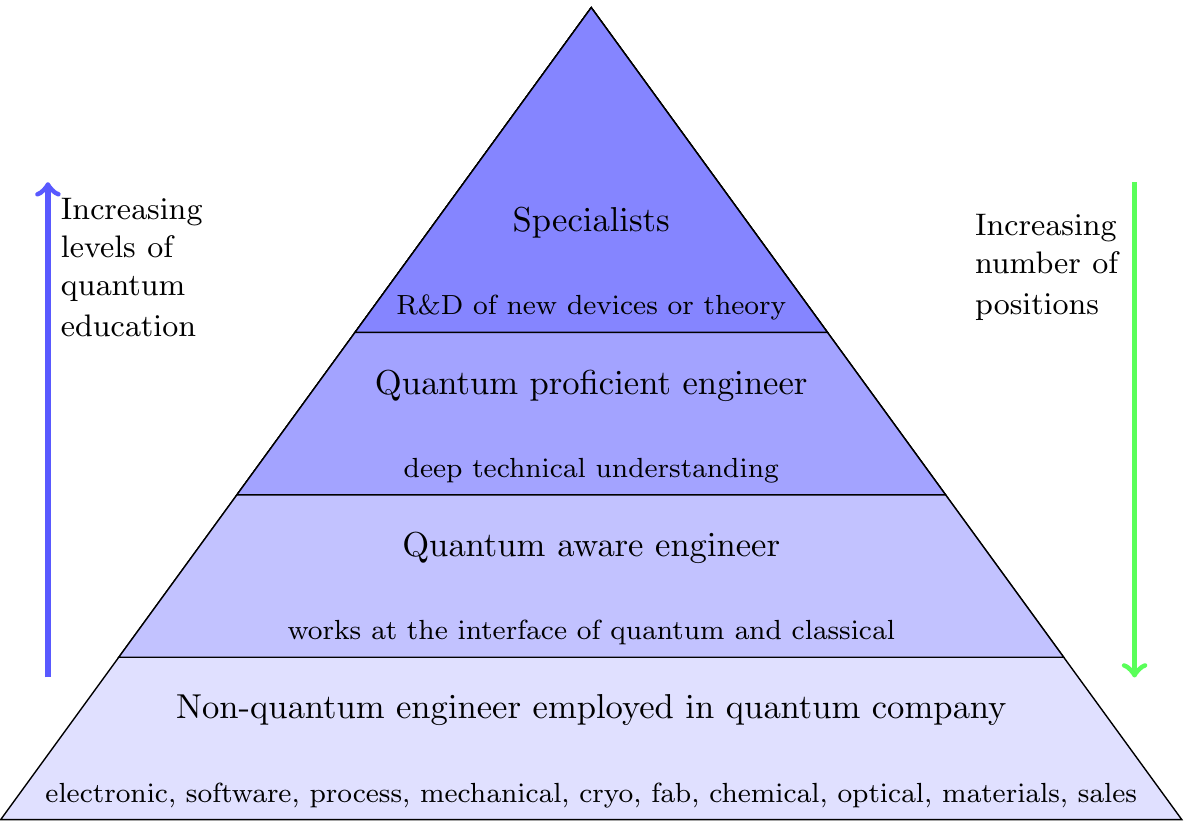}\vspace{-4pt}
          \caption{Representation of the relative number of anticipated positions for various sectors of the quantum job market and the requisite level of quantum education for each.}\label{fig:jobs} \vspace{-3pt}
    \end{center}
\end{figure*}

The purpose of this section is to identify the issues associated with quantum engineering program development and to outline possible approaches that can be tailored to individual institutions, including course development and a minor or track. In addition to resource constraints and opportunities particular to each educational institution, it is useful to keep in mind the needs of the QISE industry as it stands today and in the near future. This is shown in Fig.~\ref{fig:jobs}.  Higher levels of specialization are at the top, while lower levels of specialization, but also more jobs, are at the bottom. Positions near the very top are most likely to be filled by PhD graduates, while MS graduates will be at the middle and lower levels, and BS/BE graduates will form the base. Undergraduate program development must concern itself with filling all three of these niches. In the following, we discuss STEM education research in QISE, developing concepts-focused and advanced undergraduate courses, and practical plans for minors and tracks, closing the section with a few comments on a future quantum engineering major.\\

\subsection{QISE Education Research}
\label{ssec:education_research}

There are several pedagogical challenges associated with developing a quantum engineering program. We must contemplate the content of individual courses and degree programs. We must also bridge the gap between what we think we are teaching and what students are actually learning. Methods to do so include evidence-based active-engagement pedagogies and curricula. This bridge should ideally be research-validated in the context of STEM education, as well as be effective for students from diverse backgrounds and prior preparations. The interdisciplinary nature of QISE education implies that the diversity in students' prior preparation and background is likely to be greater in QISE-focused courses. This fact makes it even more critical to use STEM education research-validated curricula and pedagogies that focus on helping all students, not just the best-prepared, to learn. As the quantum engineering education roadmap contained in this article develops new programs and courses from scratch, we have the unusual opportunity to do things well from the ground up, rather than improving existing courses as  e.g. in quantum physics education research~\cite{mcdermott1999resource,singh2001student,mckagan2008developing,singh2008student,carr2009graduate,mckagan2010design,baily2010teaching,singh2015review}.

Development and implementation of these types of pedagogies and curricula entails thinking carefully about the learning objectives and goals of each course and aligning these with instructional design and assessment (e.g., is a pen and paper exam able to assess students' proficiency in aligning an optical system?) It also entails having a good understanding of students' prior knowledge and skills that can be built on, the common difficulties students have after traditional lecture-based instruction~\cite{difficulty1,difficulty2}, and consideration of how to leverage the diverse prior preparation of students effectively. For example, education research has studied students working in small groups on collaborative group problem solving, tutorials and clicker questions~\cite{quilt,graduate,devore,sga,ejp,aip,passante,manogue} using approaches in which individual accountability has been integrated with positive interdependence, e.g., through grade incentives. These methods have been shown to improve learning outcomes for all students. Furthermore, it is critical to contemplate how different courses build on each other in a degree program in order to maximize their benefit for students who take those courses simultaneously or sequentially.

Explicit effort should be made to ensure an equitable and inclusive learning environment, as discussed in Section~\ref{sec:diversity}, so that students from diverse demographics and backgrounds have an opportunity to excel. Moreover, validated assessment tools need to be designed to measure growth in students' knowledge and skills, as well as development in their motivational beliefs about quantum. Assessing and improving the motivational beliefs of students from different demographic groups about QISE (e.g., their QISE-related self-efficacy or sense of belonging in classes) is especially important to ensure that students from underrepresented groups also have high self-efficacy and sense of belonging, since these beliefs can impact student outcome, as well as their short and long-term retention within the field.
Along with issues of diversity, equity, and inclusion,  consideration of social, societal, ethical and sustainability issues of QISE would be beneficial, in line with directions in engineering education worldwide~\cite{rugarcia2000future,shuman2002future,goldberg2014whole}.

Finally, although there has been some education research on the effectiveness of QISE courses, more needs to be done as new programs are developed. This research needs to examine not just theory courses, but also hands-on experimental experiences and lab courses, as many of the desired skills are best learned in these environments~\cite{wilcox2017developing}.  Hands-on learning is covered in Sec.~\ref{sec:hands-on}.

\begin{figure*}[!ht]
\centering
\includegraphics[width=1.75\columnwidth]{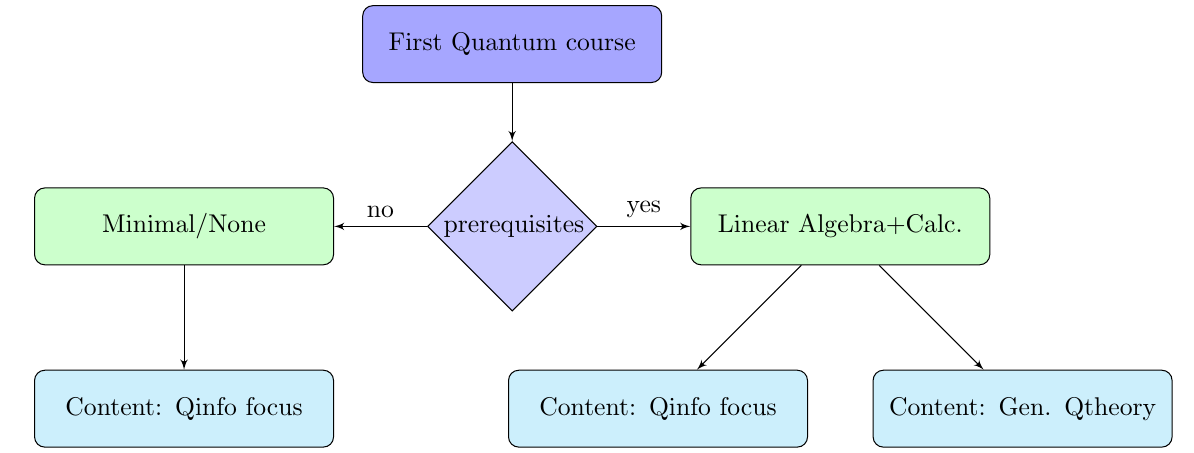}
\vspace{-5pt}
\caption{Several possibilities exist for first quantum courses for students outside physics departments. The fundamental decision is whether prerequisites are required or not. The general consensus in the community is without prerequisites the course should be focused on quantum information topics, labeled here as ``Qinfo focus''. For areas of engineering that require a quantum education, departments have a decision to make. The first option is to have students take a specialized quantum information course, resulting in quantum aware engineers. The second option is to give students a holistic quantum education, which will better prepare students for more advanced quantum courses and applications of quantum theory encountered in industrial settings, resulting in quantum proficient engineers -- see also Fig.~\ref{fig:jobs}. }\label{fig:subjects}
\end{figure*}

\subsection{Freshman-Level Concepts-Focused QISE courses: Quantum 101}
\label{ssec:concepts}

Many in the community assert that offering QISE courses at the freshman or sophomore level, without necessarily requiring linear algebra, is desirable in order to stimulate students’ interest through concepts and applications, and to help resolve structural inequities in STEM pedagogy and improve diversity, see e.g.~\cref{fig:subjects}.  For instance, such a course can be offered in community college and military school settings, and as a ``Quantum 101'' option for students who have a general interest in learning the quantum information perspective. Depending on the setting, the modules above would need adjustment or supplemented to effectively provide an entry point into QISE.

By learning concepts and applications first, without the need for advanced mathematical formalism, students may gain appreciation for the topic and intuition for the connections between concepts and applications. Moreover, avoiding advanced mathematical prerequisites or co-requisites lowers the barrier to entry for students who come to the field with less mathematical preparation, or relatively late in their education.  A similar approach is common in computer science and math departments, where programming concepts, or proof concepts courses, are often required entry points for a major or minor.  Such courses also function as recruiting tools due to their wide accessibility.

This approach in addition allows the connection between end-use applications and discussion of potential career pathways, which may appeal to technology-oriented students. Therefore, such courses may broaden the on-ramp for quantum engineers and help to recruit and retain a more diverse, more equitable, and more inclusive cohort of students into the discipline.

A potential concern with a QISE course that does not require sophisticated math is that it would require sacrificing rigor or accuracy. Surprisingly, that does not have to be the case, and there now exists a formalism that can explain quantum states, the concept of superposition, entanglement, and unitary transformations, as well as quantum algorithms rigorously without the need for linear algebra~\cite{styer2000strange,Rudolph-book,economou2020teaching,freericks2018}. The only requirement is knowledge of basic arithmetic. Through this method, which some of us have used for outreach to high school students,  in courses at the freshman level, and for courses drawing broadly on all STEM students with no quantum background, students can predict the outcome of quantum circuits. This is a nice complement to using online cloud processors or simulators, especially the drag-and-drop option that IBM offers to build circuits (which does not require text-based programming skills), as they can verify the results of their calculations using the online interface~\cite{economou2020teaching}. At this time, more research is needed to understand these approaches and their overall efficacy as courses and as bridges into more advanced material.

Finding space for such a concepts course in an engineering degree is a difficult task, but one that may be an essential entry point to the field at many institutions.  In programs where linear algebra is not a significant barrier for the student body, a first course as described in Section \ref{sec:firstCourse} can provide an efficient grounding in the concepts of quantum information, couched in the same mathematical language typically used within the field.  A one-semester introduction at this level enables students to engage with a variety of further literature and instructional resources available in the field, and can even provide enough quantum awareness to prepare engineering graduates for entry-level employment in quantum industry.  In contrast, students in many programs do find linear algebra to be a significant barrier to enrollment in a first quantum course.  In that case, a QIS concepts course can give students a solid understanding of the fundamental concepts and applications, as well as the motivation to pursue further studies.  Students who go on may then need to take more courses overall to arrive at a given level of literacy in the field, but this tradeoff can be more than worthwhile in exposing more students to the possibilities of QISE.

There are three major institutional uses for this kind of course.  First, it can be taught in community colleges, military schools, and universities that do not have the resources to create a more advanced QISE curriculum, let alone QISE degrees, to provide quantum awareness to their students. Second, it is valuable to universities that do not yet have advanced QISE courses and degrees, as a stepping stone toward building quantum engineering programs. Third, it can ease entry into more advanced and demanding courses, and provide intuition into QISE concepts without conflating them with the mathematical formalism itself.

\subsection{Considerations in Creating QISE Courses}
\label{ssec:considerations}

To develop a robust quantum workforce, it is necessary to educate students about more than simply qubit modalities, quantum circuits, and quantum algorithms. For example, many nascent offerings at universities lack education in quantum sensing, quantum communications, the theory of quantum hardware, and lab courses on quantum hardware. Rather, the majority of offerings presently focus on quantum information and quantum algorithms. Yet, this does not at all reflect the breadth of needs in industry or academia~\cite{Fox2020}. Indeed, the primary difficulty in quantum hardware and quantum technologies (sensing, communications, computing) is understanding the hardware itself, which is also changing. For example, while developing commercial quantum algorithms is one of the grand challenges of the coming decade, building quantum computers that can run these algorithms reliably at scale is equally important. Additionally, developing interconnects for quantum networks of either sensors or devices remains a challenge. As such, within a majority of vertically integrated quantum companies, the quantum algorithms team is but a fraction of the total headcount and likely composed of PhD-level employees for the foreseeable future. Thus, overemphasizing algorithms at the expense of hardware at the undergraduate level will likely miss many employment opportunities.

In contrast, the quantum engineer requires a broad knowledge of different technologies, including atoms and ions, semiconductors, superconductors, integrated optics, as well as microwave and RF control and readout~\cite{bruzewicz2019trapped,zhang2019semiconductor,kjaergaard2020superconducting,blais2020quantum}.  At the same time, the quantum engineer can pursue different areas of specialization, including communications, cryptography, and information theory; quantum computation and classical control systems; and quantum sensing and devices.  Undergraduate QISE education to date remains almost entirely housed in physics departments focused on fundamental science in preparation for the physics PhD.  Thus, a strong advantage of a quantum engineering program is to create BS/BE level students with a general knowledge of quantum technologies and specializations.  Advanced undergraduate quantum engineering programs should seek to capitalize on this opportunity at all levels by integrating many technologies and specializations either into separate topical courses, where resources are available, or into broad survey courses.  Ideally, any quantum engineering program would offer the opportunity to learn quantum communications and cryptography, quantum sensing and devices, and quantum simulations and computing.

However, one of the challenges to augmenting existing engineering programs with a minor or a track is that these programs are already highly constrained, in part by the Accreditation Board for Engineering and Technology (ABET), as well as the need for multiple core subjects and the relative lack of electives. For example, a ``Digital Signal Processing" track versus a ``Microwave Engineering" track may differ in practice by only 3 or 4 courses. That leaves essentially 3-4 available ``slots" to make a minor or a track.  Thus, it is paramount that quantum engineering programs integrate closely with existing engineering programs, as will be laid out in detail in Secs.~\ref{ssec:minor}-\ref{ssec:track}.

Another issue of  concern, and also a major opportunity, is the lack of textbooks suitable for quantum engineering.  Quantum theory textbooks are predominately written by physicists and assume a great deal of physics background, such as Hamiltonian and Lagrangian mechanics, thermodynamics and statistical physics, etc.  While there are some exceptions~\cite{miller2008quantum,schumacher2010quantum,nielsen2011quantum}, there is a need for quantum theory textbooks for non-physics majors that provide education in general aspects of quantum theory. Even more seriously, to our knowledge, no quantum engineering textbook presently exists for learning the diversity of quantum hardware at the advanced undergraduate level---see the detailed descriptions in Sec.~\ref{sec:hands-on}. Introductory and review articles exist at a wide spectrum of levels from graduate to professional QISE researchers~\cite{bruzewicz2019trapped,zhang2019semiconductor,kjaergaard2020superconducting,blais2020quantum}, and there are even a few graduate-level textbooks with hardware components, such as~\cite{orszag2016quantum}. However, much of this material is too specialized or advanced for undergraduate courses. Morover, in some cases the state-of-the-art is changing on a rapid time scale, which presents another challenge in course design and also places a burden on instructor capacity. For specific topics, there are some excellent materials aimed at undergraduates, e.g., for superconducting devices~\cite{langford2013circuit}. To create a holistic hardware course, the instructor is forced to cobble such materials together. Similar issues arise for quantum sensing and quantum characterization, verification, and validation. The latter is particularly problematic, as a major task for a quantum engineer at the moment is to assess the performance of quantum hardware and improve designs or control strategies based on this assessment. Broadly integrating classical engineering expertise in e.g., control theory into the quantum domain, i.e., \emph{quantum} control theory, will likely occur over many years.

\subsection{The Quantum Engineering Minor}
\label{ssec:minor}

A Quantum Engineering interdepartmental minor is advantageous because it supplements and leverages existing degree programs in engineering, computer science, mathematics, and the fundamental sciences. Although minors can be highly variable from institution to institution, a typical minor requires six to seven courses: three are considered core to the minor, and the remaining courses are electives. The electives may be chosen from relevant subjects within the department hosting the minor, from other departments (assuming prerequisites requirements are met), or from a student's home department. This has two consequences. First, it opens more pathways for non-traditional students to study quantum technology. Second, it avoids overloading students with subjects outside their college.
The primary decision for the host department(s) is to determine which QISE subjects should constitute the three core courses and in which departments they are taught.  Such considerations are usually particular to each academic institution, but obvious candidates are electrical engineering, computer science, materials engineering, chemistry, and physics.

There are a few quantum engineering minor programs in the U.S. with many more to appear shortly.  We provide as an example here the quantum engineering minors at the Colorado School of Mines, at the University of Colorado Boulder, and at The Virginia Polytechnic Institute and State University (Virginia Tech), which offer an initial roadmap for creating a minor at public research universities.

These are intended as working examples, and should not be construed as the only or the best programs out there. Different institutions may structure their degrees differently and may emphasize quantum computing, communications, or sensing, depending on the expertise and interests of the local QISE faculty; likewise different hardware platforms may be emphasized depending on available resources. To help ameliorate the latter, we recommend programs partnering with national labs and nearby institutions, when possible, as well as incorporate internships in industry to maximize breadth of experience in new quantum engineers.

The Colorado School of Mines quantum engineering minor requires six courses as part of the Quantum Engineering Interdisciplinary Academic Program supported by six departments including electrical and materials engineering. The first four are linear algebra, and three of the following courses: (i) fundamentals of quantum information, (ii) quantum programming, (iii) low temperature microwave measurement, (iv) quantum many-body physics, and (v) microelectronics processing. Students can either take their remaining two courses from this list to round out their quantum education, or else any two from the quantum engineering course catalog to increase specialization. The catalog is extensive, but includes existing STEM courses such as feedback control systems, digital signal processing, semiconductor device physics and design, computational materials, and machine learning.  This minor required the creation of 4 new courses targeted at quantum engineers, namely (i)-(iv).  Course (i) is covered in Sec.~\ref{sec:firstCourse}, while (ii)-(iv) were designed with the considerations of Sec.~\ref{ssec:considerations} in mind. Because Mines is a purely STEM school, many prerequisites are not necessary to specify.

The University of Colorado Boulder quantum engineering minor requires six courses with three prerequisites: programming, calculus, and linear algebra.  Students are then required to take the following three courses: (i) foundations of quantum engineering, (ii) foundations of quantum hardware, and (iii) introduction to quantum computing for the theory track, or quantum engineering lab for the experimental track.  Flexibility is built into requirement (i):  two semesters of upper division quantum mechanics in the physics department can also satisfy the requirement.  The remaining three courses are required from a large elective list which, similar to Mines, is drawn from existing STEM courses across campus, such as microwave and RF laboratory, control systems analysis, machine learning, and solid state physics.  Thus, similar to Mines, the quantum engineering minor required the creation of four new courses, some of which had already been taught in pilot form as a proof-of-principle.

The Virginia Tech QISE minor is a joint effort among seven departments/programs across the College of Science and the College of Engineering, and it requires eight courses. Five of them are mandatory: linear algebra, an introductory (freshman/sophomore level) QISE course without any prerequisites (see Sec.~\ref{ssec:concepts}), an advanced QISE course with only a linear algebra prerequisite (modules 1-4, 6, 7) described in Sec.~\ref{sec:firstCourse}, and two programming courses that include classical programming (e.g., Python), quantum programming (using online hardware and quantum languages such as Qiskit), and use of collaborative software. The fifth course is a choice between one focusing on physical quantum platforms (an expanded version of modules 9-11 in Sec.~\ref{sec:firstCourse}) and a more advanced theoretical quantum computing course. This allows students from diverse departments to deepen their QISE knowledge irrespective of whether they have a quantum mechanics background. The remaining two courses are selected out of a long list of existing electives from six departments. The creation of this minor required three new courses (the introductory QISE course and the two programming courses).

We conclude that creating a quantum engineering minor in a STEM environment can require the addition of three or four new courses, some of which may only need adaptation from preexisting instruction.  The largest barrier to the minor is hands-on training on quantum hardware, which we treat separately in Sec.~\ref{sec:hands-on}.  These examples should not be taken as an exact or detailed plan but instead as first successful attempts that can be improved upon.  We emphasize that it is more important to teaching engineering principles and design in the context of QISE than to focus on any particular quantum technology or platform, as also reflected in the diverse technologies found in Sec.~\ref{sec:hands-on}. This is critical for providing the depth of education and agility necessary to support student success in future careers, whether or not they are quantum related.

\subsection{The Quantum Engineering Track within Engineering Majors}
\label{ssec:track}

A quantum engineering track within an existing engineering program is an alternative option to the minor. In the near term, it provides
a means to more readily design a program without the need for interdepartmental collaboration. Existing engineering programs already have many subjects that are relevant to quantum engineering. Thus, a track should not be difficult to construct within such an existing framework.
Each engineering department would be able to offer multiple customized tracks comprising a series of courses that enables students to obtain domain-specific knowledge through both ``classical'' and targeted quantum subjects.  For example, an electrical engineering or computer science department could offer a quantum software track, an algorithms track, a quantum hardware track, etc. When paired with the core courses the students have chosen to form their degree program, the track provides a supplemental, in-depth training in topics relevant to a future quantum workforce, such as low-level control of quantum hardware, quantum programming languages and paradigms, quantum compilers, and the like, while fundamentally retaining the structure of their bachelors degree.

A typical quantum engineering track in an engineering program could involve the standard engineering courses plus a two-semester sequence of quantum engineering I-II, two semesters of hands-on training on quantum hardware I-II, and two specialty courses drawn from a list of standard courses available at the university such as nanofabrication, materials science, solid state devices, machine learning, etc.  For a more software-centric track implementation or option quantum computation and algorithms can take the place of quantum hardware II.  The quantum engineering I-II sequence should cover material from Sec.~\ref{sec:firstCourse} and indeed may draw directly on such a course designed for the minor in Sec.~\ref{ssec:minor}.  Depending on the engineering discipline, additional requirements may be necessary.  For example, at the U. Chicago there is a quantum engineering track in molecular engineering, which also requires intermediate electromagnetism, a necessary component for classical control systems.
We also think it would be a good idea to add to a quantum engineering track along these lines a first year freshman-level concepts-focused course, Quantum 101.  Such a course is also a highly useful addition and recruiting tool, see Sec.~\ref{ssec:concepts}.  Likewise instrumentation or lab courses offering opportunities for the quantum engineer to develop experience on two or more technologies is an important consideration, see Sec.~\ref{sec:hands-on}.

\subsection{The Future Quantum Engineering Major}
\label{ssec:major}

The minor or track may be part of existing engineering departments, or be part of an interdisciplinary program supported by several departments, including electrical engineering, materials engineering, engineering physics, computer science, chemistry, and applied mathematics, to name one of many examples discussed here.  Such quantum engineering programs may naturally evolve to the point of seeking a complete undergraduate degree titled ``quantum engineering.''  Programs can in some cases lead to stand-alone departments, as we see in the past with nano-engineering and other engineering specializations.

Whether in a program or department, the development of a quantum engineering major is a complex undertaking that may not yet be suitable for the majority of academic institutions. It is foreseeable that such majors may more commonly emerge as quantum technologies mature over the next decade and quantum engineering develops as an engineering discipline. The current ambiguity of a quantum engineering major and what it entails could adversely impact the employment prospects of students graduating with such a quantum engineering degree. For example, employers would be understandably uncertain about the preparation and training of a quantum engineer. On the other hand, a track record of hires that were electrical engineers or software engineers, now with a minor or a track in quantum engineering, may make evaluating such a candidate relatively straightforward.

\section{Promoting Diversity in Quantum Engineering Undergraduate Programs}
\label{sec:diversity}

Advances in QISE and the development of associated technologies rely on expertise from multiple disciplines including, but not limited to, applied mathematics, chemistry, computer science, electrical engineering, material science and materials engineering, and physics. These disciplines have historically struggled to be inclusive and equitable, as reflected in persistently low numbers of graduates identifying as coming from educationally marginalized racial and ethnic groups, sexual preferences, and genders, including women, the largest marginalized group of all. For example, in the U.S., the overall percentage of bachelors degrees in physics awarded to women ($20\%$), Hispanic students ($9\%$) and African American students ($4\%$) has remained stubbornly low for decades~\cite{AIPStats}, with similar trends in the other fields. Not surprisingly, the existing quantum workforce in industry is correspondingly lacking in diversity. The issues of each discipline with respect to broadening participation and facilitating success of marginalized students  (e.g., for physics~\cite{TEAMUP2020}) compound to paint a bleak picture for building diverse, equitable, and inclusive quantum engineering undergraduate programs going forward.

It is therefore imperative that curriculum designers, researchers, and university administrators implementing QISE programs think critically about issues of diversity, equity, and inclusion from the beginning. While the field and the associated undergraduate academic programs are in the early stages, we have an opportunity to effect long-lasting change in QISE and its related disciplines, and offer equitable outcomes for students from all backgrounds. Moreover, a large percentage of engineering students in the US are foreign born. A careful exploration of barriers to learning for foreign born students will be helpful toward producing the best quantum engineers.  Some of these barriers interface with diversity, equity, and inclusion issues.  We recognize that there is not a one-size-fits all solution and that education programs must be tailored to different institutions, departments, and disciplines.

\subsection{Recommendations for Course and Program Design \label{ssec:DEIgeneralrecs}}

Here, we provide some recommendations for any institution looking to add more QISE content to their curriculum in a way that also intentionally promotes diversity, equity, and inclusion. This is a world-wide issue, but the focus of this and following subsections of Sec.~\ref{sec:diversity} is on the U.S. context, needs, and initiatives.

All courses and curricula should have learning outcomes for required QISE knowledge and skills~\cite{quilt,graduate,devore,sga,ejp,aip,passante,manogue}. Courses should \emph{also} have explicit outcomes to promote a high sense of belonging, self-efficacy, and identity as a person who can excel in QISE for all students, but particularly students from historically marginalized groups, e.g., women and racial/ethnic minorities~\cite{TALR}. Self-efficacy, or belief in one's ability generally, is known to be a key predictor of success in STEM fields~\cite{quan2016connecting, litton2018increasing,marshman}. Outcomes can be evaluated through entry and exit surveys of students in each introductory QISE class. For example, in an introductory quantum mechanics class, student self-efficacy in performing quantum calculations and understanding quantum concepts should increase by the end of the course. In order to achieve these learning outcomes, it is critical for faculty to ensure that the learning environment in their courses and labs are equitable and inclusive. To this end, faculty need to be trained in inclusive mentoring approaches, such as being genuinely invested in the success of the student and having a growth mindset about their student's potential to excel~\cite{kram1985improving, Murrell1999, STEMMMentoring2020, Singh2021}. There is also evidence that brief interventions in the classroom at the beginning of the term can make the learning environment more equitable and inclusive~\cite{Binning2020}, which can have long-term effects on the success of  students from marginalized  groups~\cite{Brady2020,women_singh2021}.

Evaluation of new courses and degree programs should include key elements related to diversity, and be coupled to larger longitudinal studies of climate,  culture, and industry hiring (for example, this could be included in continuing industry surveys carried out by the Quantum Economic Development Consortium). In all data gathering efforts, it is critical to disaggregate quantitative data about student outcomes by race/ethnicity \emph{and} gender. Additionally, qualitative interviews are key to understand the impact of curricular and mentoring changes on students' experiences and persistence to degree. Recruitment of more marginalized students is not enough to achieve a truly diverse program and workforce; rather, the key metric of progress should be these students thriving in the program as well as their degree attainment and subsequent employment. In other words, \emph{equity of outcomes} is a key metric of success for any program.

One concrete recommendation for increasing diversity is to restructure science and engineering programs to accommodate QISE knowledge earlier in the curriculum. Departments should prioritize creation of a single ``quantum awareness'' introductory course as a short-term goal, see Sec.~\ref{ssec:concepts}. This is critical for institutions that are not able to implement a full QISE program. Early introduction of QISE content, the field's impact, and future career opportunities are especially important for retaining students from economically disadvantaged backgrounds (who are often also marginalized students) because it enables them to see viable career paths to well paying jobs in the quantum industry. In addition, professional development skills should be integrated into the curriculum early to facilitate student confidence. Students may also benefit from targeted training on how to work in diverse teams~\cite{Murrell1999}.

Another way to incorporate knowledge of real-world QISE applications early on is through undergraduate research experiences. Perhaps more than any other intervention, undergraduate research has been shown to increase student self-efficacy and persistence to degree, both in the general student population~\cite{Seymour2004, quan2016connecting} and for marginalized students in particular~\cite{litton2018increasing}. There is also evidence that undergraduate research experiences can encourage marginalized students to pursue higher education after their undergraduate degree, which would help to diversify the existing PhD pipeline~\cite{carpi2017cultivating}. Undergraduate research experiences can be offered in a variety of ways; for example, research could be completed at the student's home institution or in partnership with more well-resourced institutions in the area (for example, through summer programs). However, it must also be stated that not all undergraduate research experiences are equal in quality and effectiveness. High-quality mentoring is essential to achieving increased self-efficacy among students. Mentors need to be prepared to provide advice not only on research, but on other professional skills such as time-management and scientific communication~\cite{ litton2018increasing}. Increasing social support for undergraduate researchers through designated cohorts can help them build community with their peers and see themselves as engineers and scientists, something that is often difficult for marginalized students who do not see themselves reflected in the celebrated leaders of the field. Finally, there is some evidence that longer research experiences can be better for students, because they are able to form a stronger relationship with their mentor and other students~\cite{carpi2017cultivating}. To that end, we recommend that departments seriously consider implementing multi-year research programs for undergraduates during the academic year, if possible. Smaller institutions might pursue partnerships with research groups in industry or at nearby academic institutions to facilitate longer-term undergraduate research experiences. It is also important to note that there are significant barriers for many students to participate in undergraduate research, which can be magnified for many students from marginalized groups. These barriers include not having the ability to participate during the summer due to financial or family reasons. Thus, when developing new programs, it is important to consider these barriers and how to lower them so that more marginalized students can participate.

\subsection{Opportunities at Minority Serving Institutions \label{MSI}}

Historically black colleges and universities (HBCUs), Hispanic serving institutions (HSIs), tribal colleges, and native-serving institutions in the US play a significant role in promoting racial and gender diversity in engineering~\cite{MSIs2019}.  For example, from 2001 to 2009, HBCUs consistently produced over $45\%$ of all black engineering undergraduates at US institutions~\cite{Owens2012}; yet only 15 of the 107 HBCUs have ABET-accredited programs as of 2021. Further, HBCUs are known to produce large percentages ($\geq 40\%$) of black undergraduate degrees in other QISE-related disciplines contributing to diversity in most subfields (e.g., HBCUs produced an average of 25\% of CS undergraduates from 2001-2009~\cite{Owens2012}). Here, we present evidenced-based strategies for developing new opportunities in quantum engineering at minority-serving institutions through curriculum development, increasing participation in QISE and motivating engineering student success at HBCUs, HSI, tribal colleges, native-serving intuitions, and community colleges in the US.

In 1942, 20 years after the first quantum revolution, Dr. Herman Branson produced two works focused on the training of black physics students and the need for a physics-aware workforce in the context of World War II~\cite{branson1942role}. He identified the need for qualified faculty trained in physics at HBCUs,  and the development of new programs in physics at HBCUs. Today, although physics majors are being produced at an all time high in the US, the AIP TEAM UP Report the Time is Now highlighted the success of HBCUs in producing African-American physicists despite their persistent under-representation in physics in US institutions overall~\cite{TEAMUP2020}.

Similar to Branson's findings in 1942, there is now a need to train a new diverse, workforce in the context of the quantum information revolution. Establishing new programs centered around quantum engineering at HBCUs can help achieve the overall goals of the U.S. National Quantum Initiative (NQI).  Part of the larger strategy of the NQI is  prioritizing the development of education and research activities through the establishment of collaborative research and education centers across the U.S. Recent examples exist of both U.S. government and industry-led efforts to direct resources to HBCUs where the majority of black undergraduates attain degrees in STEM~\cite{IBMHBCU}. Whether industry-led or supported solely by government, new programs should utilize the best practices of Sec.~\ref{ssec:DEIgeneralrecs} when engaging diverse communities in the context of addressing issues of belonging, providing research opportunities to students, and collaborating and engaging with HBCU faculty.

Hispanic Serving Institutions (HSIs) are institutions with at least $25\%$ Hispanic undergraduate students. As of 2014, $13\%$ of post-secondary institutions were classified as HSI, but enrolled $62\%$ of undergraduate Hispanic students~\cite{Garcia2017}. Furthermore, the Hispanic population is the fastest growing major racial/ethnic group in the United States, which suggests that the role of HSIs in training the STEM workforce will only increase as time goes on~\cite{MSIs2019, TEAMUP2020}. This demographic trend indicates that implementing new quantum engineering programs at HSIs now could meaningfully increase the participation of Hispanic Americans in the quantum workforce in the long-term. Some HSIs are also research-intensive institutions, but many are smaller, primarily undergraduate-serving institutions~\cite{Garcia2017}. There are opportunities for partnerships between HSIs with and without engineering programs, enabling sharing of resources and curricula for introductory QISE courses between the two. Industry and government initiatives similar to those for HBCUs should also be considered for HSIs. Following this guidance, new programs in quantum engineering can expect to find that each subfield of QISE also benefits with respect to their overall diversity and equity efforts.

\subsection{Transfer Pathways from Two-Year and Four-Year Institutions}

Many engineering programs have connections with two-year community colleges and four-year institutions without engineering programs, and these partnerships provide an important pathway for students to enter engineering professions. In 2000, as many as $40\%$ of students who received a bachelor's or master's degree in engineering attended a community college at some point~\cite{EnhancingCC2006}. Several studies have shown that transfer students are equally or more successful compared to non-transfer students in completing degrees in engineering~\cite{cosentino2014black, berhane2019transfer, winberg2018persistence}.  Thus this student group is key to growing the quantum workforce. New QISE programs should build connections with existing partnerships where possible, for example, by helping community colleges implement an introduction to quantum science course (see Secs.~\ref{sec:firstCourse} and~\ref{ssec:concepts}) or by offering summer internships for students. Building new partnerships with community colleges serving a large minority population should be prioritized~\cite{EnhancingCC2006}.

Near-term opportunities abound for leveraging curriculum development efforts in QISE to create bridges between a more diverse cohort of students and careers in quantum engineering. For example, the introductory modules discussed in Sec.~\ref{sec:firstCourse} intended for everyone (E) should be accessible to students at 2- and 4-year institutions with existing transfer pathways to established STEM programs. This could be realized as either reserved seats in classes offered at the destination institution with existing cross-registration agreements or offering the class at the transfer school with a guaranteed transfer credit, depending on local circumstances. If an effort is made to emphasize applications and potential career trajectories in QISE during these introductory classes, this approach may help build bridges from students unfamiliar with the STEM landscape to degree programs and careers.

\subsection{Industry's Role in Promoting Diversity in Undergraduate Quantum Engineering}

External stakeholders should provide powerful incentives to promote diversity, equity, and inclusion in undergraduate quantum engineering programs, which in turn will help to diversify the future workforce. It is first important to recognize that increasing diversity in industry promotes diversity in undergraduate education, and vice versa. In this section, we highlight some ways that industry can work together with academia to expand and diversify the quantum workforce.

Students from marginalized  groups are more likely to apply for a particular major if they see their peers from that major being hired by industry~\cite{national2019minority}. Presently, the majority of employees at quantum technology companies have a PhD in physics or engineering, but it is unlikely that the PhD qualification will be needed for many of the positions required for a thriving quantum industry~\cite{Fox2020}. Industry can support the twin goals of workforce growth \emph{and} diversification in several ways.

First, extra effort both by universities and industry is needed with respect to placing marginalized student  in industry internships. This is because a strong predictor of whether a student will be hired by a company is if the student has completed an internship with the company, as explored e.g. in~\cite{knouse2008benefits,sanahuja2015effects}. If internships are too limited, or not available at an undergraduate level, alternative programs such as participating in the open-source community (e.g. contributing to GitHub repositories or participating in hackathons) to get real-world coding experience may also improve chances of placement~\cite{national2019minority}.

Second, industry must strive to democratize access to online training and resources (including quantum computing access) for quantum engineers, and in particular focus on partnerships with MSIs, as discussed in Sec.~\ref{MSI}. In fact, this is already occurring in some places, for example through the IBM strategic partnership with San Jos\'{e} State University~\cite{IBMSJSU}, the IBM-HBCU Quantum Center~\cite{IBMHBCU}, and Google's related diversity and open source initiatives~\cite{googleDiversity}.

Lastly, industry and academic institutions need to continue benchmarking progress on improving diversity. For example, today in the U.S., the Quantum Economic Development Consortium already has the participation of companies to perform workforce surveys, so we recommend it expand its workforce development charter to track diversity metrics as well. The U.S. National Science Foundation has grant programs that can similarly evaluate student outcomes of academic institutions, and professional societies such as the American Society for Engineering Education (ASEE), American Institute of Physics (AIP), American Physical Society (APS), the American Chemical Society (ACS), etc. already play a large role in obtaining such statistics. Implementing periodic evaluation of academic and industry diversity initiatives is essential to ensure that programs meet their desired outcomes. Similar efforts can be undertaken in many nations throughout the world.

\subsection{Summary of Diversity, Equity, and Inclusion Recommendations}

Our recommendations fall into three categories: climate; curriculum and program evaluation; and industry.

First, regarding the climate for diversity, we recommend faculty and industry research mentors be given training in best practices for mentoring, including building authenticity and trust with students and engaging in culturally aware communication that builds on students' strengths, since this is  especially critical for the success of marginalized students~\cite{kram1985improving, Murrell1999, STEMMMentoring2020, Singh2021}.  Such training programs are available at many colleges and universities and where they are lacking can be leveraged from partner institutions. Departments developing QISE courses and programs should conduct periodic climate surveys of students and make them publicly available, to establish new best practices and continually evaluate what is and is not working.  It is also very helpful to integrate high-quality undergraduate research experiences early (see Sec.~\ref{sec:hands-on}). Longer, multi-semester research experiences are preferable to short ones if possible~\cite{carpi2017cultivating}.

Second, regarding curriculum and program evaluation, we emphasize that at minimum a quantum awareness or concepts course be offered to introduce students to the field early, at the freshman or sophomore level. This is particularly important for institutions that do not offer engineering degrees but which have many students that transfer to engineering programs. This may also be applicable to other QISE fields such as physics and computer science. In program evaluation, we recommend disaggregating all data on outcomes by gender \emph{and} race/ethnicity.  Instructors should consider implementing brief interventions in the classroom early in the semester to make the learning environment more inclusive and equitable~\cite{Binning2020}. Course and departmental goals around equity should focus on \emph{equity of outcomes} for students, e.g., ensuring that marginalized students thrive in the program and, after completing their degree, have successful careers in QISE. A singular focus on recruitment should be avoided~\cite{Singh2021}.  We recommend professional development (writing and culturally aware communication skills) be integrated into the required curriculum~\cite{Murrell1999}. Quantum programs should restructure how and when the first year of quantum science is taught. Frontloading applications could make career trajectories more transparent, see Sec.~\ref{ssec:concepts}.  It is important to engage 2-year and 4-year institutions without engineering programs to open up transfer pathways~\cite{cosentino2014black, berhane2019transfer, winberg2018persistence}.

Finally, regarding industry, we recommend that students be provided equitable access to industry educational resources and internships, e.g., through formal partnerships with minority serving institutions. It is important to evaluate existing workforce needs and long-term goals and to communicate those needs to degree programs and curriculum designers. Industry can additionally develop credentialing pathways for marginalized students through internships and/or open-source educational materials.

\section{Hands-on Training on Quantum Hardware}
\label{sec:hands-on}

Hands-on learning in QISE is key to creating viable quantum engineers. Experimental research in QISE is extremely diverse; the tools used in quantum optics labs differ greatly from those used in microwave-controlled superconducting quantum circuits, for example. However, one of the chief opportunities of quantum information as a lens to view these experiments is its ability to link them with a common language. Educational labs that make this connection physical, as well as mathematical, by showing that similar experimental conclusions arise from experiments that look completely different, will help to develop a cross-disciplinary skill set and allow for a quantum workforce that can communicate across traditional hardware barriers.

Assembling and selecting hands-on labs for a quantum engineering course or program will not be a one-size-fits-all solution. Each program will have constraints, including budgets and faculty expertise. While plug-and-play resources in quantum engineering are available in some cases, many are expensive, while lower-cost do-it-yourself approaches require significant expertise. Affordable hands-on training availability may depend on cross-institution exchange of expertise.  For example, the Advanced Laboratory Physics Association (ALPhA) currently provides funds and structure for faculty members to exchange hands-on training on advanced undergraduate physics lab modules~\cite{alpha}; support for similar efforts in quantum engineering could be transformative for many fledgling programs.

The diversity of hardware modalities makes it impractical to effectively cover the breadth and depth of quantum systems that form the basis for current research and industrial R\&D. Effective training strategies should therefore seek to cover a core but limited collection of required knowledge, supplemented by a selected subset of more in-depth studies of particular platforms. The three major areas of QISE, communication, computing, and sensing, all involve preparation, measurement, and control of quantum states. Core components of hands-on training should therefore include an exploration of preparation, measurement,  and control, while highlighting how quantum approaches differ from the classical techniques that are routinely covered in introductory engineering laboratory courses. Even if specific institutional barriers restrict hands-on lab work to a single platform, or if most of the interactivity comes from remote resources, adding simpler demonstrations of alternative platforms will provide students larger context through which to understand the field. This section will provide examples of affordable approaches to covering these core capabilities.

Below we include a wide variety of platforms. It is impossible to predict which platforms currently under investigation, or yet to be discovered, may end up being the best subject(s) for future quantum engineers. Therefore, inline with best practices in engineering education across engineering disciplines, it is important to train quantum engineers in general principles and design, not to be a technician on one particular technology.  For this reason, we emphasize that programs should try to cover at least two platforms as part of the hands-on courses.

\subsection{Optics}

Quantum optics in the visible regime provides a relatively budget-friendly toolbox for lab activities, allowing students to gain hands-on experience with quantum state preparation, manipulation, and measurement. Many experiments can be performed using the polarization state of a laser beam as an analogy to a true quantum state, such as single-qubit state and channel tomography. It is possible to use classical laser beams to simulate quantum key distribution (QKD) in a way that has lingering security loopholes, but gives students an interactive and tactile project making explicit use of many quantum-relevant features, such as superposition and measurement disturbance~\cite{utama2020hands}. Student experiments with classical laser beams can also provide students with an essential toolbox of classical skills necessary in experimental quantum optics, such as optical fiber alignment and detection electronics.

For experiments that require quantum states of light, such as nonlocality experiments, spontaneous parametric down-conversion of visible or near-ultraviolet laser light in a nonlinear optical crystal is a well-developed and affordable approach. There are many articles describing how to set up spontaneous parametric down-conversion sources and related accessible optical experiments in quantum state preparation and measurement for students~\cite{Galvez2005,Pearson2010}. The equipment needed is also available as a plug-and-play system from multiple companies, with costs currently on the order of \$20k USD~\cite{SPDCkits}.\footnote{Mention of commercial suppliers is provided for information only and is not an endorsement of the products of a particular company.}

Whether quantum or classically simulated, optics experiments clearly show how changes of the quantum state impact measurement results. This is most simply accomplished through the polarization degree of freedom, where polarizers are used for projective measurement and combinations of half- and quarter-wave plates can be used to rotate states and measurements to arbitrary points on the Bloch sphere. However, the same effects can also be shown in space using Mach-Zehnder interferometers and time through Franson interferometry, although some effort will be required for stabilization in these degrees of freedom. Partial coupling between degrees of freedom, such as a birefringent crystal that couples time and polarization, can be used to simulate decoherence.

With quantum systems that feature single photons such as those generated by spontaneous parametric down-conversion, students may show Bell nonlocality through violations of the Clauser-Horne-Shimony-Holt inequality, non-classical photon correlation functions through Hanbury-Brown-Twiss interferometry, Wheeler’s delayed-choice quantum eraser, photon bunching in a Hong-Ou-Mandel interferometer, multi-qubit state tomography, entanglement-enabled quantum key distribution in the Ekert protocol, determination of one- and two-photon coherence times, and more~\cite{beck2012quantum}. Many protocols relevant to quantum information, such as the bulk optical C-NOT implementation and GHZ state creation, are also possible to implement, although they require specially tailored photon pair sources and/or multi-pair emissions, which may be prohibitively expensive or alignment-sensitive for lab courses.

\subsection{Atoms and Ions}

Other possibilities for experiments with individual quanta can be contemplated with nitrogen vacancy center, neutral atom, or trapped ion hardware. Nitrogen vacancy centers, in particular, are well suited for sensing applications and demonstration kits are commercially available~\cite{NVkits}. Complete demonstration kits at the level of individual quantum operations based on the other hardware platforms are not readily available. Compact and cost effective hardware for laser cooling and magneto-optical trapping of atoms can be purchased~\cite{Atomkits}. Besides laser cooling, this type of hardware can be used for demonstrating quantum state control by optical pumping. The extension to experiments with single quanta using optically trapped atoms or electromagnetically trapped ions still requires substantial local expertise and infrastructure. Partnerships with industry could enable such laboratory experiences.

Ultracold atoms provide another setting where quantum phenomena including superposition, interference, and tunneling can be observed. The apparatus to produce Bose Einstein condensates is complex, comparable to a dilution refrigerator in cost, and not practical to maintain without locally available expertise~\cite{simpleBEC}. An alternative is cloud access to a commercial machine that can be remotely operated~\cite{CQAlbert}.

\subsection{Cryogenic and Solid State}

Quantum solid-state platforms such as superconducting circuits and quantum dots require cryogenic operation at a temperature of 4K or lower, while experiments deep in the quantum regime require dilution refrigerators to reach temperatures down to 10 mK. Such equipment is highly specialized and expensive (\$300k is the entry level cost for a dilution refrigerator). Thus, deployment of dilution refrigerator experiments in an undergraduate laboratory will rely on local expertise or grouping the cost across regional schools. Similar to trapped ions and atoms, institutional partnerships could be essential to bring experience with such techniques within the grasp of students at a broader range of colleges and universities.

Higher-temperature alternatives include implementing other types of cryogenic systems such as the Quantum Design Physical Property Measurement System (PPMS)\footnote{\url{https://qd-europe.com/at/en/product/physical-property-measurement-system-ppms/}} which can cool to 1.9K and thus would facilitate labs based on ancillary measurements such as superconducting transition temperatures or basic microwave resonator transmission data acquisition.

As an alternative that will help
undergraduates gain familiarity with control of spins, techniques for manipulating and measuring them, and the terminology of coherence, experience with a pulsed NMR apparatus can be valuable.  Complete  setups are commercially available~\cite{pulsedNMR}.

\subsection{Nanofabrication}

Nanofabrication is crucial to the creation of many types of quantum devices, including in photonic quantum computing, superconducting quantum computing, spin qubits, and ion traps. A laboratory or set of laboratories focusing on the design and implementation of a fabrication process for a simple device such as a lumped element resonator or Josephson junction would give students an understanding of the requirements of device fabrication, knowledge which is much needed in many implementations of quantum computing. Partnerships with nanofabrication facilities or groups with specialized equipment for qubit device design would augment student learning in this area.

\subsection{The Quantum-Classical Interface}
All quantum technology requires a means of passing information between the classical and quantum domain, for instance, in the readout and control of qubits or quantum sensing devices. The quantum-classical interface (QCI) is a catch-all term for the electronic  and optical sub-systems  of  readout  and  control,  such  as data  converters,  amplifiers,  signal  sources,  and  digital  logic responsible  for  generating  and  detecting  readout and control waveforms, sequencing and synchronising them, as well as the infrastructure that connects those signal paths to the physical quantum  devices  that  encode  quantum  information.  Such infrastructure   comprises   cabling,   packaging, optical fibres,  chip-interconnects, resonators,  and  on-chip  routing  and  multiplexing  approaches that   together,   constitute   IO   management between the classical and quantum worlds.

The QCI provides an important opportunity  for hands-on training, making use of the classical software and hardware needed to support and enable quantum experiments. Laboratory courses featuring microwave engineering, programming of embedded systems, signal generators, digitizers, and related hardware can provide skills of broad applicability for experimental R\&D. Its worth noting that in the context of teaching, reasonably sophisticated electronics can be sourced for minimal cost in comparison to other domains of quantum hardware. Further, such electronic sub-systems provide an ideal platform for the development of generic problem solving skills such as trouble-shooting and debugging.

Modern communication systems and the engineering framework for their design provide a solid foundation for developing quantum control and readout platforms. Examples include modulation and demodulation techniques, non-reciprocal elements, noise mitigation and approaches to bandwidth narrowing, for instance, using lock-in amplifiers for detecting weak signals. With respect to these topics, there is much common ground between electronic and optical or photonic systems. Indeed, radio frequency and microwave circuits (the basis for qubit control and readout) mix electrical and optical concepts and terminology. An ability to map and bridge these domains is a particularly useful attribute of the quantum engineer.

Moreover, the challenges associated with the quantum-classical interface are likely to be major hurdles for the scale-up of quantum computers and quantum networks. The complexity of these systems over the next decade will rival the most sophisticated technological platforms ever constructed. Beyond the challenge of the hardware itself, quantum engineers must also simultaneously be able to work at various levels of abstraction, bridge fields, and leverage long-forgotten knowledge with new research discoveries.

\subsection{Tools and Involvement from Industry}

Undergraduate engineering education has a long history of using tools from industry in the classroom. For QISE education, providing the physical hardware to university students affordably and at scale is a challenge. However, within the quantum computing industry, innovations drawing from the infrastructure of classical computing have led to increased access to quantum computing resources. While not strictly hands-on in nature, these tools provide an opportunity for students to experience authentic quantum devices. These quantum computing tools can be categorized into either open-source software or hardware access via the cloud. Open source software packages such as QisKit~\cite{QiskitGettingStarted},  Cirq~\cite{CirqGettingStarted}, and Katas~\cite{KatasGettingStarted} are essentially free for students and educators to use, and can be used to simulate quantum computers up to 30+ qubits on an affordable classical computer accessible to most undergraduate students. Furthermore, higher level libraries such as TensorFlow Quantum~\cite{TFQ}, OpenFermion~\cite{OpenFermion}, and QisKit Aqua~\cite{QiskitAqua} can increase accessibility to students who are already familiar with machine learning or chemistry simulations.

For hardware access, companies including AWS, Google, IBM, IonQ, and Rigetti have made available some of their quantum processors for use via the cloud by the public and in the classroom. For example, IBM provides free access to their smaller systems, and Google has implemented batch execution of student assignments on their cloud systems.  Although computer explorations are not a substitute for hands-on experimentation, for institutions that are not able to provide a laboratory experience, there are widely available and valuable educational tools supported by major companies in the areas of quantum computing and simulation. Companies including Google, IBM, and Microsoft provide extensive online tutorial material as well as quantum circuit simulators, among other publicly available interfaces. For example, IBM's QisKit platform provides free access to students wishing to experiment with quantum circuit design and in addition to simulation tools allows users to run examples on real hardware via cloud access.

In addition to the above industry tools, internships ~\cite{GoogleInternships} ~\cite{IBMInternships} not only provide hands-on opportunities, but can also provide the education and experience of working directly within industrial settings, which all have very different cultures and goals compared to academic labs.

\section{Summary and Key Recommendations}
\label{sec:recommendations}

The rapid expansion of quantum information science and engineering (QISE) outside of the research lab and in industrial applications necessitates growth of a diverse workforce with increasing quantum knowledge and skills. Development of QISE applications including communication, computing, and sensing require people at all levels from K-12 to the PhD who are trained in quantum-related science and engineering. One strong focus of new training is at the undergraduate level for engineers. To facilitate the development and implementation of new quantum engineering education opportunities, we present an initial roadmap for those creating these new programs, including suggested courses and modules, approaches to engage students in QISE training, and ways to rethink and create diverse, inclusive and equitable education.  Our recommendations will enable bachelor's level engineers to achieve two levels of QISE training, quantum aware and quantum proficient.

Below, we outline the broad recommendations to consider when developing new quantum engineering education programs at the undergraduate level.

\begin{itemize}

   \item Traditionally, quantum courses and programs have been contained within physics departments. To prepare engineers for jobs in the quantum industry, new programs and training should be created in engineering departments with collaborations from science and math departments.

    \item To create quantum-aware engineers, we have detailed, module by module, the development of a first QISE course for STEM students that can be implemented in many different academic environments.  Such a course could be adjusted for different contexts, with additional modules, and would be sufficient for any engineer to obtain the minimum quantum expertise needed to participate in the QISE industry.

   \item To create quantum-proficient engineers at a higher level than just being quantum aware, and at the current stage of the quantum industry, we recommend universities and colleges develop new minors in quantum engineering or tracks embedded in traditional majors, rather than full undergraduate degree programs. As the quantum industry grows over the next decade, full undergraduate degrees in quantum engineering may be desired and can be natural extensions of  minor and track programs.

    \item Minors or tracks in quantum engineering can be offered at many colleges and universities, as we suggest a minimum of only three or four new courses need to be created, with additional electives drawing from standard STEM course offerings.

    \item We suggest a QISE course accessible to non-STEM students can be taught using very little math and instead focus on basic concepts and applications---Quantum Information Science and Engineering 101. This course can recruit students into a minor program, onboard students into the minor, and serve as general QISE education accessible to all STEM students from freshman year on. The focus on applications could make career trajectories more transparent to students as well. We recommend this type of course could be implemented broadly, including at community colleges and military schools, which could facilitate students' transition to a 4-year institution.

    \item One important component to any minor program is hands-on experimental training, as many of the jobs in the quantum industry require this expertise. We recommend a variety of hardware platforms where students can get this experience. We note that less expensive ``classical'' options, or partnerships with institutions with more resources, could help students at institutions with less infrastructure gain hands-on experience. Additionally, we recommend integration of high-quality undergraduate research experiences early in students' academic careers, with longer, multi-semester research experiences being preferable as this will also improve diversity.

    \item It is important to make sure these new courses and programs are effective at helping students to achieve the learning goals for the courses and programs. Toward that end, we recommend continued and expanded STEM education research be done for QISE, especially the engineering context, to establish effective practices in this new domain.

    \item As we begin to develop new programs, we have the opportunity to focus on creating a more diverse, inclusive, and equitable environment for our students. We have several recommendations to help achieve these goals. Departments developing QISE courses and programs should conduct periodic climate surveys of students and make the results publicly available. This will help to establish effective practices and provide formative feedback to improve the programs.  Aligned with this, we suggest that course and program goals around equity should focus on \emph{equity of outcomes} for students, i.e., degree attainment and employment, and avoid a singular focus on recruitment~\cite{Singh2021}.

\end{itemize}


Our recommendations, although ultimately reflecting only the authors, draw heavily on community input. As QISE engineering programs develop and mature based on ongoing education research other programs in QISE fields will benefit from lessons learned. In particular, the``Quantum 101" course, if implemented, could provide data on how conceptual quantum science and technology can be taught in different settings, potentially providing an opportunity to modify current curricula within other QISE-related departments. Currently, there is a breadth of quantum physics education research that could be leveraged~\cite{mcdermott1999resource,singh2001student,mckagan2008developing,singh2008student,carr2009graduate,mckagan2010design,baily2010teaching,singh2015review} but more must be done in the context of engineering fields.  Additionally, because QISE is cross-disciplinary, additional QISE curricula and education research across a range of settings could inform course design and pedagogy within QISE fields outside of engineering.

We acknowledge the extensive thoughts and feedback from the QISE community developed in a series of documents in the February 2021 NSF Workshop on Quantum Engineering Education with 480 quantum information scientists and engineers in attendance from across academia, government, industry, and national labs.  We also acknowledge useful conversations with Abida Mukarram on the specific needs of U.S. community colleges.


\begin{thebibliography}{100}
\bibitem{SPDCkits}
Complete demonstration kits for SPDC experiments are available from Qubitekk,
  Qutools, and Thorlabs and among others.
\bibitem{NVkits}
Demonstration kits for experiments with NV centers are available from Qutools
  and CIQTEK.
\bibitem{Atomkits}
Compact setups for laser cooling of Rb atoms are available from ColdQuanta.
\bibitem{CQAlbert}
ColdQuanta provides cloud access to a quantum matter machine that is remotely
  programmable through a graphical user interface.
\bibitem{pulsedNMR}
Advanced undergraduate pulsed NMR setups are available from TeachSpin.
\end{thebibliography}

\end{document}